\documentclass[sigconf,nonacm]{acmart}
\usepackage{graphicx}
\newcommand{\utility}{u}
\newcommand{\noisedUtility}{\hat{u}}
\newcommand{\wtp}{w}
\newcommand{\price}{p}
\newcommand{\user}{i}
\newcommand{\itm}{j}
\newcommand{\noise}{\epsilon}
\newcommand{\dimension}{d}
\newcommand{\DecisionSet}{D}

\usepackage{tikz}
\usetikzlibrary{bayesnet}
\usetikzlibrary{arrows}
\usepackage{color}
\usepackage{graphicx}
\usepackage{caption}
\usepackage{subcaption}
\usepackage{dirtytalk}
\usepackage{multirow}
\usepackage{paralist}
\renewcommand\footnotetextcopyrightpermission[1]{}

\usepackage{subfiles} 
\begin{document}

\title{Welfare-Optimized Recommender Systems}

\author{Benjamin Heymann}
\email{b.heymann@criteo.com}
\affiliation{%
  \institution{Criteo Technology, équipe-projet commune FAIRPLAY}
  \streetaddress{32 Blanche}
  \city{Paris}\country{France}
}

\author{Flavian Vasile}
\email{f.vasile@criteo.com}
\affiliation{%
  \institution{Criteo AI Lab}
  \streetaddress{32 Blanche}
  \city{Paris}
  \country{France}}

\author{David Rohde}
\email{d.rohde@criteo.com}
\affiliation{%
  \institution{Criteo Criteo AI Lab}
  \city{Paris}
\country{France}
}

\begin{abstract}
We present a recommender system  based on the Random Utility Model. 
Online shoppers are modeled as rational decision makers with limited information, and the   recommendation task  is formulated as the problem of optimally  enriching  the shopper's awareness set.
Notably, the price information   and the shopper's Willingness-To-Pay  play  crucial roles.
Furthermore,  to better account for the  commercial nature of the recommendation, we unify the retailer and shoppers'  contradictory objectives  into a single welfare metric, which we propose as a new recommendation goal.
 We test our framework on  synthetic data and show its performance in a wide range of scenarios.
This new framework, that was absent from the Recommender System literature, opens the door to Welfare-Optimized Recommender Systems,  couponing,  and price optimization. \end{abstract}

\ccsdesc[500]{Information systems~Recommender systems}

\keywords{discrete choice, random utility models, neural networks, recommendation, dynamic pricing, welfare optimization}

\maketitle

\section{Introduction}
\subsection{Recommender Systems}

\begin{figure}[h]
\includegraphics[width=0.35\textwidth]{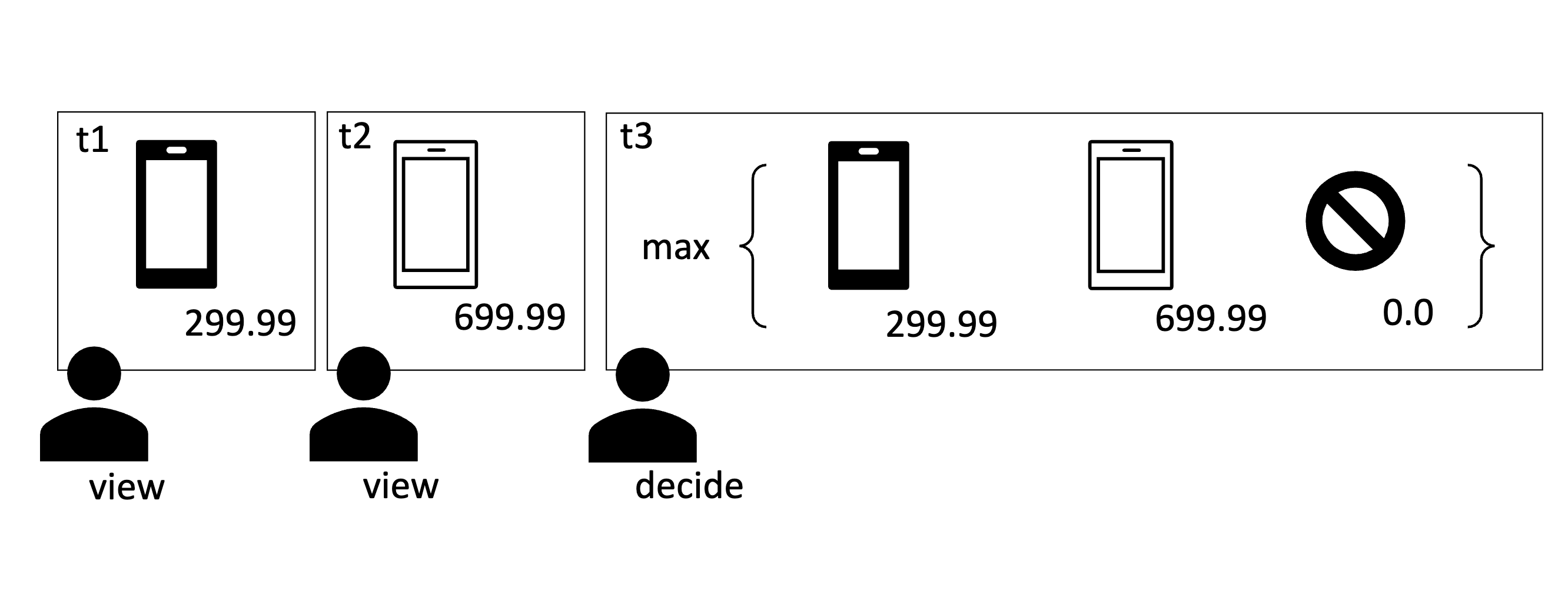}
\caption{The model of the user as a rational decision-maker with a limited awareness set.
At time t1 and t2, the user observes different models of phones, with their prices. 
At time t3, they decide whether to buy one of the two phones, or to leave the website empty-handed. This decision is the output of a rational process of utility maximization that takes into account both the features and the prices of the two phones. The novelty of our framework is that we make explicit the difference between what the user knows (the awareness set) and the entire catalog.}
\label{fig:graphical_story}
\end{figure}

Recommender Systems (RS) have proven successful almost everywhere in the digital economy, as they are a crucial component of the engines that run digital advertising,  e-commerce websites, digital media, music and video providers and social networks.
For the last few years, there has been a growing effort --- partially inspired by economic theory--- in providing a principled way to design these systems.
We contribute to this research effort by providing a proof of concept that recommender systems used by e-tailer and advertisers could be welfare maximizers.
To do so, we leverage a user behavioural model summarized in Figure~\ref{fig:graphical_story}, and  which will be formalized in Section ~\ref{sec:shopperbehavioralmodel}.

\subsection{Shortcomings of current approaches}
Most reward-optimized recommendation systems measure an abstract form of user utility and not an actual monetary value. This situation likely stems from the preponderance of clicks as immediate reward feedback in real-world systems. But as the field and the industry mature, we need consistent and rigorous approaches for conversion-optimized recommendation systems.  The apparent similarity of conversions and clicks from the point of the merchant/advertiser is misleading, and the current click-optimization approaches are conceptually lacking when it comes to conversion modelling. 
The three main differences  are:
\begin{itemize}
    \item While clicks are free for the users, conversions cost real money, and therefore the item price should play a big role in the utility computation within the Recommender System. In other words, the price of an item is a disincentive to buy, but not to click.
    \item Due to finite budgets, users have to make choices and items serving the same need are mainly in competition in the user shopping process. This is much better modelled as a Categorical distribution over the choices, instead of the standard Bernoulli used for click modelling.
    \item While clicks are conditioned on existing recommendation systems policies (they are bandit feedback by nature), conversions are in a large part organic and therefore less sensitive to changes in recommendation policy.
\end{itemize}

Also, because
the goal of recommendation is to facilitate two-party transactions, 
one needs  to clearly state on whose behalf  the Recommender System is operating (the buyer or the seller's side).

\subsection{Our approach}
\begin{figure}[h]
\includegraphics[width=0.35\textwidth]{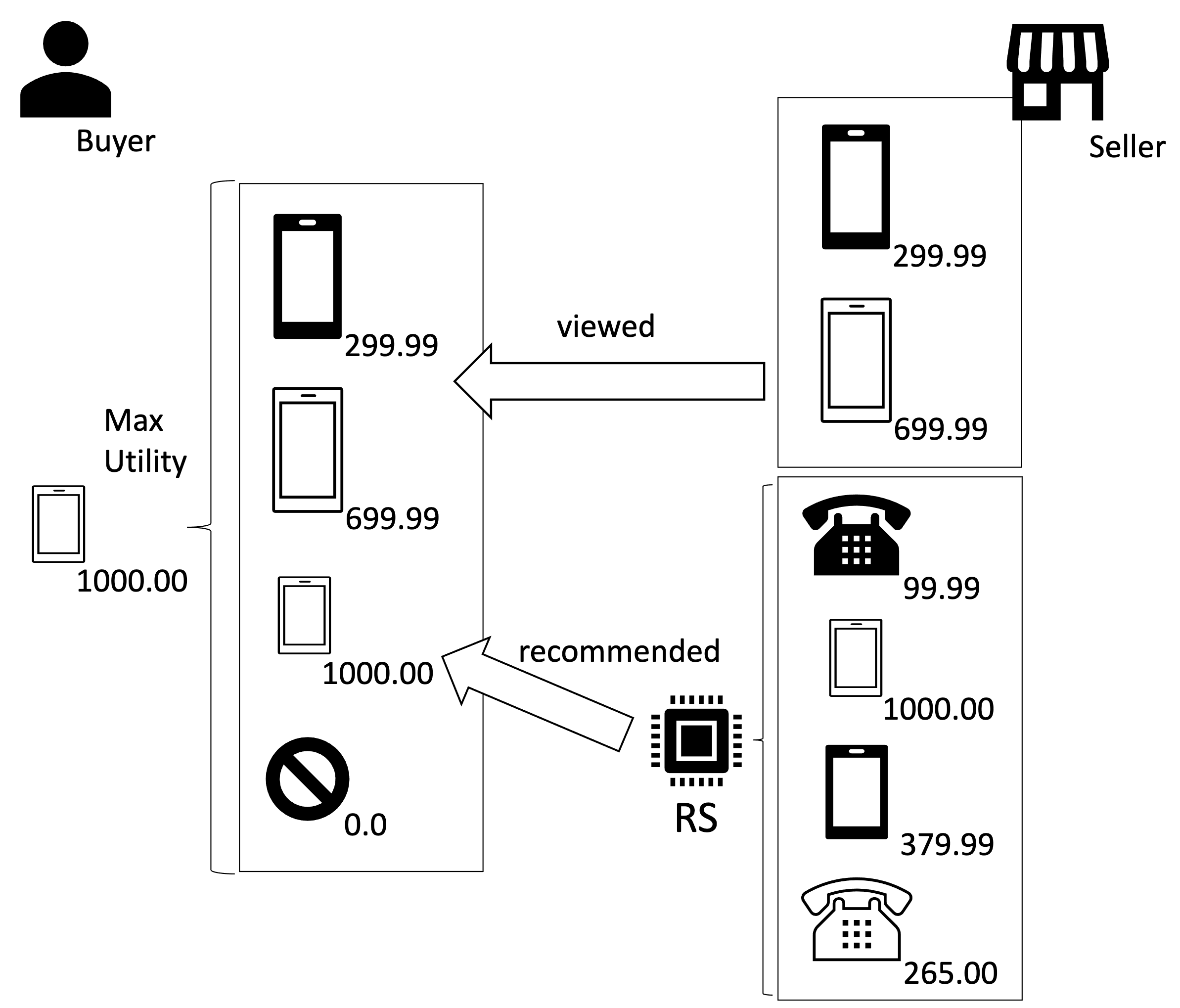}
\caption{The role of recommendation in a limited information setup. Based on its knowledge of the user, the RS points the user to a new item that is likely to be useful. The value of  RS is  equal to the difference in utility between the user's decisions with or without the recommendation.}
\label{fig:reco_value}
\end{figure}

 To address all these shortcomings, 
 we rely at learning time on the user's browsing history
  to infer the set of items over which the user is making the decision to buy or to leave ( Figure~\ref{fig:graphical_story}), and then 
 we frame the problem of recommendation as the problem of finding \textit{the optimal enrichment of the user's awareness set} (Figure \ref{fig:reco_value}).

Also, \emph{we propose a novel recommendation objective: welfare maximization}, moving the focus from ways of splitting the pie --- which is the main focus of multi-stakeholder literature--- to how to maximize the pie. 
This leads us to an objective based on the Willingness-To-Pay (WTP), which measures the maximum price that the user is willing to pay for an item.  More precisely, it is the price point where the user is indifferent to buying or not the item~\cite{savage1971elicitation}. 
To be able to maximize the WTP objective, we propose a simple algorithm which can be seen as an extension of the classical Matrix Factorization (MF) algorithm~\cite{xue2017deep,5197422} that incorporates price and that can serve both as a user decision model and as a WTP estimator.

\subsection{Paper contents}
In Section~\ref{sec:related-work}, we cover the related work on recommendation with prices.
In Section~\ref{sec:pareto} we discuss potential applications of the Welfare objective for Recommendation. 
In Section~\ref{sec:shopperbehavioralmodel}, we introduce the user discrete choice model, and propose RUM-MF (random utility model via matrix factorization), a simple extension of existing MF algorithms. In Section~\ref{sec:experiments}, we introduce our simulated environment and present our results against  
multiple performance-optimizing baselines  on classical and welfare-based metrics. In Section~\ref{sec:conclusion}, we conclude and outline future research directions.

\section{Related Work}
\label{sec:related-work}

\subsection{Performance-optimized Recommendation}

Performance-optimized Recommendation has a long history, with many initial works  related to the problems of  click-through rate (\textit{CTR}) optimization for online advertising and search ranking. There are two categories of methods. The first   relies on the label or reward given by the user, the second   directly learns an order on the items.

The first class of methods was motivated by  online advertising, and mostly relies on likelihood-based models. They are preferred due to their calibration properties, and range from the use of large-scale logistic regression with sparse features as in \cite{chapelle2014simple}, Gradient Boosted Trees \cite{dave2010learning}, Factorization and Field-aware Factorization Machines \cite{rendle2010factorization,juan2016field}, and more recently, Deep Neural Networks  \cite{swaminathan2015counterfactual}, Bayesian Methods \cite{sakhi2020blob}, Variational Autoencoders \cite{liang2018variational}, and Policy-based methods \cite{swaminathan2015counterfactual}, leading to state-of-the-art results.

The second class of methods was motivated by search engine design and mostly relies on learning-to-rank  models. This includes classical ranking methods such as SVM-Rank \cite{joachims2002optimizing}, Triplet Loss and Syamese Networks \cite{dong2018triplet} and that have been more recently supplemented by Bayesian Personalized Ranking (BPR) \cite{rendle2012bpr,joachims2017unbiased}.

\subsection{Price in Recommendation}
Moving to specialized approaches for conversion modelling and the use of price in recommendation, 
the literature is quite varied and ranges from full user shopping decision modelling via multiple objective matrix factorization of the shopping data as in \cite{wan2017modeling},
to ranking approaches  that supplements user - item affinity measures with global and local budget preferences, as in \cite{hu_learning_2018}. 
More recently, \cite{zheng2020price} proposes a Graph Convolution Network-based method to infer category~-~specific price elasticity,  and~\cite{wadhwa2020personalizing} explores several methods to compute item price bands and explicitly model user price affinity for various types of products.%

\subsection{Recommendation and Econometric Models}
Probably the closest area of research is the existing body of work around the use of economic-inspired recommendation models.
In \cite{zhang_economic_2016}, the authors outline the relationship between recommendation and surplus or welfare maximization and propose a way to estimate the personalized WTP based on the concavity of the utility function. 
In our paper, we extend their ideas to support what we think is the realistic case of limited awareness, and propose a different way to estimate the WTP that is closer to standard Matrix Factorization approaches. Similar limitations of the user decision set where already noted in the economic theory~\cite{LLERAS201770} literature.

In  \cite{zhao_e-commerce_2015},  the authors propose learning the WTP with auctions on mechanical turk, and using it for personalized promotion. In the follow-up paper \cite{zhao_multi-product_2017}, the authors propose Multiproduct Utility Maximization (MPUM) based on the complementarity property of some existing choices. In \cite{hu_learning_2018} the authors propose learning the in-session WTP using views and purchase decisions.
More recently, in a series of papers \cite{ge_maximizing_2019, zhichao_xu_e-commerce_2020} the authors propose two economic recommendation objectives, Marginal Utility per Dollar and  Weighted Expected Utility. Unlike in our approach, the authors are only considering the user-side. 
On the vendor side, \cite{das2009maximizing, jannach_price_2017} both raise the point that the product prices have an impact on the seller's revenue and profit and propose alternative ways to incorporate this information in the final recommendation ranking function. In \cite{lu2014show}, the authors propose an extension to dynamic recommendation strategies that take into account demand saturation, and show that the problem is NP-complete and propose tractable heuristics.

\subsection{Random Utility Models}
 Discrete Choice Theory is a classical branch of Economics, and Random Utility Models are one of the main tools of the discipline. 
Discrete Choice theory has its origins in the 60s with work such as \cite{arrow1961capital}. Its use to answer marketing questions was pioneered by Daniel McFadden in \cite{mcfadden1981econometric}. He  won a Nobel Prize for his work on Random Utility Models (RUM), and more precisely on Conditional Logit models \cite{mcfadden1973conditional}.
Random Utility models (RUM) have been introduced in the 1970s, a time where the availability of user consumer data was scarce and the observability of the salient user and product features was reduced. However, in modern times, a lot of the user shopping activity moved online, therefore tremendously increasing the volume of available user shopping data and its granularity. Furthermore, it can be argued that in many cases, the modeller has access to a full view of the product information available to the consumer at the moment of the decision. That means that at learning time, we have total access to the full feature set used in the decision-making process, albeit in a raw form.
For this reason, recent work has explored the possibility of applying RUM to consumer basket modelling and dynamic pricing (SHOPPER) \cite{donnelly2019counterfactual}. However, though the authors drew parallels to existing recommendation models, the model was not used and benchmarked for recommendation.
\section{ Welfare and Recommendation}
\label{sec:pareto}

There are many ways to frame the problem of designing a Recommender System algorithm as a machine learning problem.
For some, it is the question of guessing the most probable next item, or of matrix completion, for some others,  it is an exploration-exploitation  problem (bandit setting).
A more recent line of approaches frames the problem of designing a Recommender System algorithm in terms of economic value maximization \cite{zhang_economic_2016}. 
Our claim is that in the presence of sales data, recommendation algorithms can use the price information to directly optimize the welfare of the whole system (which is  made of the advertisers and the users), instead of maximizing the probability of an action (click or conversion for instance). 
We use the simple but powerful idea that users convert when the price is below a stochastic threshold value, the Willingness-To-Pay (WTP)~\cite{savage1971elicitation}, and  leverage  a mature recommendation algorithm --Matrix Factorization (MF)-- to learn the user decision process and the underlying WTP.

We concur with authors in ~\cite{abdollahpouri2019beyond} who wrote that 
\say{recommender systems serve multiple goals and that the purely user-centered approach found in most academic research does not allow all such goals to
enter into their design and evaluation}.
In the case of online shopping, if the Recommender System is a user assistant, one has to raise the concern of contradictory objectives because on the one hand, the sellers might want to maximize their revenues or profit, and, on the other hand, the user is looking to maximize their utility.

The difference between what the user actually pays and the WTP can be seen as a surplus/utility that the user wants to maximize.
On the seller side, we suppose the seller to be a revenue maximizer -- we could equivalently suppose a profit maximizing criterion by taking into account the seller's procurement cost. 
\emph{We propose to  maximize the total welfare of the system buyer+seller, defined as the sum of the utility of the buyer and the seller.} If, when a sale occurs, the buyer's utility is their surplus (WTP - price) and the seller utility is the revenue  generated by the sale (the price), then the sum of the two results in the system maximizing the expected WTP. 

\section{A shopper's behavioural model}
\label{sec:shopperbehavioralmodel}
\subsection{User Decision Model}

In this section, we present a simple model for the user behaviour. 
Such model can be understood by describing what happens between the moment the user reaches the e-tailer page, to the moment the user leaves or decides to buy something. 
In this description, we suppose that the user is looking for  one specific category of product (for example, a pair of shoes or a mobile phone).
By design, our model does not encompass the purchase of several goods, hence while not adapted for grocery shopping, we believe this encompasses a lot of the situations where  Recommender Systems play a key role as  shopping assistants.

Our main assumption is that while the user browses  the e-tailer website --- for example, looking for a new mobile phone ---, they \textit{learn} about the different alternatives proposed by the website. 
More precisely,\emph{we suppose that we can infer the awareness set over which the user decides to purchase something}, or simply leave. 
For instance, in order to build the awareness set, one can take the browsing history and add the most popular items in the category, as well as the previous purchases. 

By doing so, and assuming the user is rational and risk-neutral, we can define its decision process as choosing the item $\itm^\star$ that maximizes their utility over the associated awareness set at time t $\DecisionSet_\user(t)$ , to which we add the \textbf{no buy} option (by definition, the no buy option brings 0 utility). More formally, we have that:
\begin{align}
   \itm^\star(t)\in \mbox{argmax}_{\itm\in \DecisionSet_\user(t)} \utility_{\user,\itm}(t),
\end{align}
where $\utility_{\user,\itm}(t)$ is 
 the total utility of  user $\user$ for item $\itm$:
 \begin{align}
\label{eq:utility}
    \utility_{\user,\itm}(t)=\wtp_{\user,\itm}-\price_{\itm}(t),
\end{align}
 computed as the surplus between the Willingness-To-Pay (WTP) $\wtp_{\user,\itm}$ and the price $\price_{\itm}(t)$ of item $\itm$ at time $t$ . In our case, time $t$ is the index of the browsing session, to support the case where the user has performed multiple shopping trips.

\subsection{Random Utility Maximization/ Matrix Factorization (RUM-MF)}
Next, we introduce two additional assumptions that allow us to link the utility maximization to  Matrix Factorization. 

\subsubsection{Gumbel noise}
We extend the Equation~\eqref{eq:utility} and  connect our model with Random Utility Models: we suppose that there exists noise $\noise_{\user,\itm}(t)$ on the latent variable $\utility_{\user,\itm}$,  such that at each decision time, the shopper makes their decision based on
$    \noisedUtility_{\user,\itm}(t)=\utility_{\user,\itm}+\kappa_\user\noise_{\user,\itm}(t),
$,
where $\kappa_\user$ is a positive, scalar model parameter. 
With this change,  the shopper's decision --- knowing $\user$ and $\price$ ---, denoted $x_{\user}(t)$ is now random. 
Also, we suppose the noise samples $\noise_{\user,\itm}(t)$ to be independent and follow a standard Gumbel distribution, which leads us to the classical softmax decision rule, for $\itm\in\DecisionSet_\user(t)$:
\begin{align}
    \Pr(x_{\user}(t)=\itm) =\frac{ \exp(\utility_{\user,\itm}(t)/\kappa_\user)}{\sum_{\itm'\in \DecisionSet_\user(t)} \exp(\utility_{\user,\itm'}(t)/\kappa_\user)}.
\end{align}
\subsubsection{Factorizable form}
We suppose there exists a dimension $\dimension$, and a d-dimensional vectorial representation of the users $X_{\user}$ and of the items $Y_{\itm}$ such that 
$
    \wtp_{\user,\itm} = X_{\user}\cdot Y_{\itm},
$
where $\cdot$ is the dot product. 
Such representation is very close to the Matrix Factorization, and can benefit from pre-existing user and/or item embeddings, that can be built for example using clicks, number of co-occurrences, user segmentation or catalog input.
With these two assumptions, the decision model becomes
\begin{align}
\label{eq:rum_decision}
    \Pr(x_{\user}(t)=\itm) =\frac{ \exp((X_\user\cdot Y_\itm-\price_\itm(t))/\kappa_\user)}{\sum_{\itm'\in \DecisionSet_\user(t)} \exp((X_\user\cdot Y_{\itm'} -\price_{\itm'}(t))/\kappa_\user)}.
\end{align}

It is classical but notable that in our formulation, we get a relation between the utility, and the odds ratio between two alternatives:
\begin{align}
\label{eq:odd}
  \log   \frac{\Pr(x_{\user}(t)=\itm|(\itm,\itm') \in  D_{\user}(t))}{\Pr(x_{\user}(t)=\itm'|(\itm,\itm') \in  D_{\user}(t))}= \frac{\utility_{\user,\itm}-\utility_{\user,\itm'}}{\kappa_\user}.
\end{align}
The interpretation of Equation~\ref{eq:odd} is  that  the  odd ratio between two alternatives is fully explained by the  difference in utility the two alternatives provide to the user. 
So if $\kappa_\user$ is known,  we can relate  the WTP, the prices, and the odd ratio between two alternatives.

\section{Experiments}
\label{sec:experiments}

 While many RecSys performance studies rely on datasets with logged partial feedback to back their claim, such methodology is not adapted to the problem at hand, since we cannot have access to the true WTP in real life datasets. For this reason, we build our own simulated environment in which we observe the behavior of our proposed method against existing state-of-the-art. For more discussion on the use of simulators in the analysis of Recommender Systems, see ~\cite{ekstrand2021simurec}.
 For space consideration, we only report here the main insights. 
 The experiments detail as well as the tables of results can be found in the long version of this article as well as in the appendix, that contains an extended version of this section. 

Our data generator takes as input
the number of users and items, the number of sessions per user, the number of items seen in each session and the dimension  $d$ of the latent space. 
It then generates two vectors from a $d$-dimensional Gaussian distribution. 
Those two vectors are then used as  means for generating the users and items $d$-dimensional representations (by sampling from a Gaussian distribution). 
Each item's price is set to be  revenue maximizing  plus uniform noise. Once the items and the users are generated, we generate user sessions by exposing users to random sub-samples of items from the catalog and at the end of each session we run the utility computation to create the session outcome.
We implemented  our models in PyTorch and used Adam \cite{kingma2014adam} to optimize over a $L_2$-regularized cross-entropy loss. 

To compute our metrics, we assume that our recommendations will be shown in slates of various sizes. We use a greedy approach to fill the banner using the top k products ranked by their expected Value per Sale (eVPS):
$
   eVPS_{j|i} = \Pr(\hat{u}_{i,j} > 0) \times Value_j
$
where $Value_j$ is the value of the sale, which can be one of the four: volume (value equal to 1), user's utility, seller's revenue and overall welfare. While the greedy slate-filling approach is not optimal, it is by far the most popular approach and it has been shown in \cite{derakhshan2022product} to be a good approximation of the global maximum.
To note, because of the greedy strategy, metrics that are not aligned with the user's utility, such as the seller's revenue,  might not be monotonous in $k$.

We tested the following methods.
\begin{itemize}
\item \emph{Oracle} Represents the best k products for each user, according to the true item features and true user preferences. We run experiments where we rank the products by their user's utility (\emph{Oracle-utility}) or by the system welfare (\emph{Oracle-welfare}).
\item \emph{BestOf} Returns the k most popular products in the training data, sorted by number of sales.
\item \emph{RUM-MF} Returns the k most likely to be bought products according to the proposed model.
\item \emph{MF-SM(Softmax)} This is the same method as \emph{RUM-MF}, but without using the price in the model training. As a note, we predict the WTP $\wtp_\user$ by considering that the innerproduct between the user and item vectors is still approximating the utility
$u_\user$ for the pair and adding the item price to it:
$
    \Pr(x_{\user}(t)=\itm)_{MF-SM} =\frac{ \exp(X_\user\cdot Y_\itm)}{\sum_{\itm'\in \DecisionSet_\user(t)} \exp(X_\user\cdot Y_{\itm'})}.
$
\item \emph{MF-PCLICK} Treats every conversion as an independent outcome, similar to Factorization Machine methods used for Click-through Rate Prediction \cite{juan2016field}. Since the method is not recovering the willingness-to-pay, only two objectives are available for the slate generation, namely volume of sales and seller's revenue: $
    \Pr(x_{\user}(t)=\itm)_{MF-PCLICK} =\sigma(X_\user\cdot Y_\itm).
$
\end{itemize}
\subsection{Findings}
Optimizing for welfare leads to the best welfare output.
As expected, optimizing for welfare works and it leads to the best welfare metric, \emph{RUM-MF-welfare}, surpassing by more than 30\% the performance matrix factorization methods (see long version for details).

We also observe that RUM-MF outperforms the other methods not only in Welfare metrics but also in classical Precision and Sales metrics, albeit by a smaller margin.  The performance is maintained for slates with k>1

Switching to our formulation can lead to improvements even in the absence of prices.
We observe that while not as competitive as \emph{RUM-MF}, \emph{MF-SM} outperforms clearly the \emph{MF-PCLICK} approach. Indeed, even in the absence of prices, the categorical formulation of the problem and the presence of the no-buy option in the model leads to a better estimation of the user's utility vector.

\section{Conclusion}
\label{sec:conclusion}

We believe that as new  algorithms tend to  leverage the price information,
welfare maximizing Recommender Systems are likely to
become a reality in the near future.  
Because they will explicitly model the user's behaviour, 
this new class of RS will allow for a better understanding of the user interaction with the RS. 
The shift from click maximizer to welfare maximizer will permit a better control of the value sharing between the different stakeholders.
Also, we  implicitly relied on the classical Luce axiom of independence from irrelevant alternative, which might not be always satisfied. 
This is, however, a general problem in discrete choice theory that is not specific to our approach, so that one could investigate if the current approach could borrow,  for instance, from  nested decision models. 
 Our goal was to provide a minimal working algorithm, but in practice such algorithm could be adapted to the available information. 
For example, if we have access to embedded representations of the user and/or the items (based on clicks or sales or browsing behaviours...), then we can  use these representations as inputs in a neural network.

\newpage

\newpage
 \appendix
 \onecolumn
\section{Experiments: extended version}
\label{sec:experiments}

 While many RecSys performance studies rely on datasets with logged partial feedback to back their claim, such methodology is not adapted to the problem at hand, since we cannot have access to the true WTP in real life datasets. For this reason, we build our own simulated environment in which we observe the behavior of our proposed method against existing state-of-the-art. For more discussion on the use of simulators in the analysis of Recommender Systems, see  ~\cite{ekstrand2021simurec}. 

\subsection{The simulator}
Our data generator takes as input
the number of users and items, the number of sessions per user, the number of items seen in each session and the dimension  $d$ of the latent space. 
It then generates two vectors from a $d$-dimensional Gaussian distribution. 
Those two vectors are then used as  means for generating the users and items $d$-dimensional representations (by sampling from a Gaussian distribution). 
Each item's price is set to be  revenue maximizing  plus uniform noise. Once the items and the users are generated, we generate user sessions by exposing users to random sub-samples of items from the catalog and at the end of each session we run the utility computation to create the session outcome.
\paragraph{Simulation environments}
For our experiments, we define three types of configurations with various levels of difficulty for our simulator, as shown in Table~\ref{tbl:configs}. We classify the level of difficulty of the environment based on the user and item-level sparsity of the training data. In the medium configurations, each user sees 30\% of the catalog and each item has associated around 300 events. In the hard case each user sees only 3\% of the catalog and each item is visited on average only 30 times. Both the item and user vectors are of dimension 10 and the associated Gaussian distributions have their variance set to 3. The only other simulator parameter is the price noise  which we kept fixed  to the range 0-5.
\begin{table}[]
\begin{tabular}{llll}
\toprule
                 & Medium1 & Medium2 & Hard \\
\midrule
nb-sessions      & 3       & 15      & 3    \\
nb-items-session & 10      & 2       & 10   \\
nb-users         & 1000    & 1000    & 1000 \\
nb-prods         & 100     & 100     & 1000 \\
dimension        & 10      & 10      & 10  \\
\bottomrule
\end{tabular}
\caption{The three environment configurations}
\label{tbl:configs}
\end{table}
\subsection{Training}
We implemented all our models in PyTorch and used Adam \cite{kingma2014adam} to optimize over a $L_2$-regularized cross-entropy loss. 
In the implementation of the \emph{RUM-MF} model we opted for a slightly different parametrization than the one in Equation.\ref{eq:rum_decision}, which is still equivalent to it, but in which the parameter $\kappa_\user$ parameter can be seen as the price sensitivity of the user specific to the product category:
\begin{align}
    \Pr(x_{\user}(t)=\itm)_{RUM-MF} =\frac{ \exp(X_\user\cdot Y_\itm- \kappa_\user \price_\itm(t))}{\sum_{\itm'\in \DecisionSet_\user(t)} \exp(X_\user\cdot Y_{\itm'} - \kappa_\user \price_{\itm'}(t))}
\end{align}

The  Neural Network jointly learns the embedding and $\kappa$ on sales signals by LLH maximization.
As a result, we will estimate the WTP as the ratio of the dot product between the user and item vectors and the user sensitivity to price:
$
    w_{i,j} = \frac{X_\user\cdot Y_\itm}{\kappa_\user}
$.
When estimating WTP, we cap all our learnt sensitivities to minimum 0.1.

\subsection{Metrics}

In order to compute our metrics, we assume that our recommendations will be shown in slates of various sizes. We use a greedy approach to fill the banner using the top k products ranked by their expected Value per Sale (eVPS), as defined below: 
\begin{align}
\label{eq:expectedvalue}
   eVPS_{j|i} = \Pr(\hat{u}_{i,j} > 0) \times Value_j
\end{align}

\noindent
where $Value_j$ is the value of the sale, which can be one of the four: volume (value equal to 1), user's utility, seller's revenue and overall welfare. While the greedy slate-filling approach is not optimal, it is by far the most popular approach and it has been shown in \cite{derakhshan2022product} to be a good approximation of the global maximum.
To note, because of the greedy strategy, metrics that are not aligned with the user's utility, such as the seller's revenue,  might not be monotonous in $k$.

To define our metrics, we introduce the following notation: 
\begin{itemize}
    \item $j^{*}_\user$ is the true optimal choice for the user $i$ over the catalog, $j^{*}_i = argmax_{j \in C} u_{i,j}$
    \item $S^k_i$ is the set containing the top k choices sorted by their predicted utility for the user i, to which we add the \textbf{no buy} option. \emph{$S^k_i$ is practically an awareness set built by the RS.}
    \item $j^*_{ik}$ is the optimal choice for the user $i$ out of the set of items present in the slate $S^k_i$, $j^*_{ik} = argmax_{j \in S^k_i} u_{i,j}$
\end{itemize}

We divide our metrics into classic metrics, such as 
Precision@k   and  Sales@k --  respectively  $|j^{*}_i \in \hat{S}^k_i|/|I|
$ and $| u_{i, j^*_{ik}} > 0 |/|I|
$-- and  novel metrics, such as 
  BuyerUtility@k,  SellerRevenue@k  and 
  Welfare@k  ($\sum_{i \in I} u_{i, j^*_{ik} }/|I|
$, 
 $\sum_{i \in I} p_{j^*_{ik}} /|I|$ and $ \sum_{i \in I} (u_{i, j^*_{ik}} + p_{j^*_{ik}}) /|I|$)
. 
For the classical metrics, while both of them are well-known, computing probability of sale at various slate sizes is novel, and only possible due to our access to an oracle. 

\subsection{Methods}
Below we present the set of methods that we take into consideration in our experiments:

\paragraph{Oracle} Represents the best k products for each user, according to the true item features and true user preferences. We run experiments where we rank the products by their user's utility (\emph{Oracle-utility}) or by the system welfare (\emph{Oracle-welfare}).
\paragraph{BestOf} Returns the k most popular products in the training data, sorted by number of sales.
\paragraph{RUM-MF} Returns the k most likely to be bought products according to the proposed model.
\paragraph{MF-SM(Softmax)} This is the same method as \emph{RUM-MF}, but without using the price in the model training. As a note, we predict the WTP $\wtp_\user$ by considering that the innerproduct between the user and item vectors is still approximating the utility
$u_\user$ for the pair and adding the item price to it:
$
    \Pr(x_{\user}(t)=\itm)_{MF-SM} =\frac{ \exp(X_\user\cdot Y_\itm)}{\sum_{\itm'\in \DecisionSet_\user(t)} \exp(X_\user\cdot Y_{\itm'})}
$

\paragraph{MF-PCLICK} Treats every conversion as an independent outcome, similar to Factorization Machine methods used for Click-through Rate Prediction \cite{juan2016field}. Since the method is not recovering the willingness-to-pay, only two objectives are available for the slate generation, namely volume of sales and seller's revenue: $
    \Pr(x_{\user}(t)=\itm)_{MF-PCLICK} =\sigma(X_\user\cdot Y_\itm)
$

\subsection{Results}

Below we summarize our main findings:

\paragraph{Optimizing for welfare leads to the best welfare output}
On the first axis, we wanted to understand the impact of different choices of objective (and the associated value per sale --- VPS) on the 5 metrics, as shown in Table 4. The first observation is that, as expected, optimizing for welfare works and it leads to the best welfare metric, \emph{RUM-MF-welfare}, surpassing by more than 30\% the performance of the \emph{MF-SM} method and almost doubling the performance of \emph{MF-PCLICK}. This observation is confirmed by Table 5, that shows that the magnitude of the improvement in performance is maintained over all configurations and runs.
 
\begin{table*}[t]
\label{tbl:objectives}
  \centering
\begin{tabular}{llrrrrr}
\toprule
  Algo & Objective &  Welfare@k &  Utility@k &  Revenue@k &  Sales@k &  Precision@k \\
\midrule
oracle &   welfare &      21.51 &      11.31 &      10.19 &     0.89 &         0.41 \\
oracle &   utility &      20.58 &      14.32 &       6.25 &     1.00 &         1.00 \\
bestof &     sales &       8.39 &       5.27 &       3.12 &     0.57 &         0.11 \\
\midrule
   rum-mf &   welfare & \textbf{11.74+/-0.26} & \textbf{6.78+/-0.21} & 4.96+/-0.11 & \textbf{0.78+/-0.01} & \textbf{0.14+/-0.01} \\
   rum-mf &   utility &  10.67+/-0.3 & \textbf{6.59+/-0.21} & 4.08+/-0.12 & 0.75+/-0.02 & 0.12+/-0.01 \\
   rum-mf &   revenue & \textbf{11.43+/-0.36} & 4.37+/-0.18 & \textbf{7.06+/-0.22} & 0.64+/-0.02 & 0.08+/-0.01 \\
   rum-mf &     sales &  10.82+/-0.3 & \textbf{6.68+/-0.21} & 4.14+/-0.11 & \textbf{0.78+/-0.01} & 0.12+/-0.01 \\
   \midrule
    mf-sm &   welfare &  8.59+/-0.25 & 5.97+/-0.21 & 2.62+/-0.06 & 0.75+/-0.01 &  0.1+/-0.01 \\
    mf-sm &   utility &  8.16+/-0.21 & 5.73+/-0.18 & 2.42+/-0.05 & 0.73+/-0.01 & 0.09+/-0.01 \\
    mf-sm &   revenue &  8.93+/-0.24 & 6.08+/-0.19 & 2.85+/-0.07 & 0.75+/-0.01 & 0.11+/-0.01 \\
    mf-sm &     sales &  7.53+/-0.22 & 5.51+/-0.18 & 2.01+/-0.05 & 0.73+/-0.01 & 0.08+/-0.01 \\
    \midrule
mf-pclick &   revenue &  7.03+/-0.21 & 5.13+/-0.18 &  1.9+/-0.05 & 0.69+/-0.01 & 0.07+/-0.01 \\
mf-pclick &     sales &   6.42+/-0.2 & 4.88+/-0.18 & 1.54+/-0.04 & 0.69+/-0.02 & 0.06+/-0.01 \\
\bottomrule
\end{tabular}
\caption{Performance of various models on top product recommendation (k=1) (Medium2, run2)}
\end{table*}

\begin{table}[]
\label{tbl:welfare}
\begin{tabular}{lllllll}
\toprule
          & \multicolumn{2}{l}{Medium1} & \multicolumn{2}{l}{Medium2} & \multicolumn{2}{l}{Hard1} \\
\midrule
          & run1         & run2         & run1         & run2         & run1        & run2        \\
\midrule
oracle    & 20.10        & 23.38        & 19.32        & 21.51        & 27.81       & 28.50       \\
bestof    & 5.75         & 9.06         & 5.30         & 8.39         & 5.83        & 5.80        \\
\midrule
mf-rum    & \textbf{11.61}        & \textbf{12.89}        & \textbf{7.88}         & \textbf{11.74}        & \textbf{7.06}        & \textbf{7.43}        \\
mf-sm     & 8.85         & 8.94         & 7.31         & 8.93         & 4.2         & 5.06        \\
mf-pclick & 7.26         & 9.06         & 5.23         & 7.03         & 5.17        & 5.61  \\
\bottomrule
\end{tabular}
\caption{Welfare@1 metric over multiple environments}
\end{table}

\paragraph{RUM-MF outperforms the baselines in all other metrics}
In the same Table 4, we can observe that RUM-MF outperforms the other methods not only in Welfare metrics but also in classical Precision and Sales metrics, albeit by a smaller margin.

\paragraph{The performance is maintained for slates with k>1}
In Table 6. we confirm that the results hold also for banners/slates with k bigger than 1. This is important since in real-world recommendations, the performance is judged not over single item recommendations but over recommendation sets. We observe that, as expected, the gap in performance becomes smaller as k becomes bigger and all methods start retrieving good items in their top k set.

\paragraph{Switching to our formulation can lead to improvements even in the absence of prices.}
We observe both in Table 4. and 5. that while not as competitive as \emph{RUM-MF}, \emph{MF-SM} outperforms clearly the \emph{MF-PCLICK} approach. Indeed, even in the absence of prices, the categorical formulation of the problem and the presence of the no-buy option in the model leads to a better estimation of the user's utility vector.

\begin{table}[]
\label{tbl:varyingk}
\begin{tabular}{llll}
\toprule
          & Welfare@1 & Welfare@5 & Welfare@10 \\
\midrule 
oracle    & 21.51     & 20.84     & 20.62      \\
bestof    & 8.39      & 11.48     & 14.9       \\
\midrule
rum-mf    & \textbf{11.74}     & \textbf{16.92}     & \textbf{18.27}      \\
mf-sm     & 8.93      & 14.51     & 16.56      \\
mf-pclick & 7.03      & 12.7      & 14.95     \\
\bottomrule
\end{tabular}
\caption{Welfare metric over multiple k (Medium2, run2)}
\end{table}

 \end{document}